\documentclass[acmlarge,screen]{acmart}
\newcommand{{\proj}}[0]{ProChatter}
\newcommand{{\callm}}[0]{CA-LLM}
\usepackage{graphicx}
\usepackage{tikz}
\usetikzlibrary{trees}
\usepackage{subfig}
\AtBeginDocument{%
  \providecommand\BibTeX{{%
    \normalfont B\kern-0.5em{\scshape i\kern-0.25em b}\kern-0.8em\TeX}}}

\setcopyright{acmcopyright}
\copyrightyear{2018}
\acmYear{2018}
\acmDOI{XXXXXXX.XXXXXXX}

\acmConference[Conference acronym 'XX]{Make sure to enter the correct
  conference title from your rights confirmation emai}{June 03--05,
  2018}{Woodstock, NY}
\acmPrice{15.00}
\acmISBN{978-1-4503-XXXX-X/18/06}

\begin{document}

\title[``Confrontation or Acceptance'']{``Confrontation or Acceptance'': Understanding Novice Visual Artists' Perception towards AI-assisted Art Creation}

\author{Shuning Zhang}
\email{Zhang.sn314@gmail.com}
\author{Shixuan Li}
\email{li-sx24@mails.tsinghua.edu.cn}
\affiliation{
  \institution{Tsinghua University}
  % \streetaddress{P.O. Box 1212}
  \city{Beijing}
  \country{China}
  \postcode{100084}
}

% \affiliation{%

% \affiliation{%

% \affiliation{%

% \affiliation{%

% \affiliation{%

% \affiliation{%

% \affiliation{%


\renewcommand{\shortauthors}{Trovato and Tobin, et al.}


\begin{abstract}
    The rise of Generative Artificial Intelligence (G-AI) has transformed the creative arts landscape by producing novel artwork, whereas in the same time raising ethical concerns. While previous studies have addressed these concerns from technical and societal viewpoints, there is a lack of discussion from an HCI perspective, especially considering the community's perception and the visual artists as human factors. Our study investigates G-AI's impact on visual artists and their relationship with GAI to inform HCI research. We conducted semi-structured interviews with 20 novice visual artists from an art college in the university with G-AI courses and practices. Our findings reveal (1) the mis-conception and the evolving adoption of visual artists, (2) the miscellaneous opinions of the society on visual artists' creative work, and (3) the co-existence of confrontation and collaboration between visual artists and G-AI. We explore future HCI research opportunities to address these issues.
\end{abstract}

\begin{CCSXML}
<ccs2012>
   <concept>
       <concept_id>10003120.10003121.10011748</concept_id>
       <concept_desc>Human-centered computing~Empirical studies in HCI</concept_desc>
       <concept_significance>300</concept_significance>
       </concept>
   <concept>
       <concept_id>10010405.10010469</concept_id>
       <concept_desc>Applied computing~Arts and humanities</concept_desc>
       <concept_significance>500</concept_significance>
       </concept>
   <concept>
       <concept_id>10003120.10003121.10003122.10003334</concept_id>
       <concept_desc>Human-centered computing~User studies</concept_desc>
       <concept_significance>300</concept_significance>
       </concept>
 </ccs2012>
\end{CCSXML}

\ccsdesc[300]{Human-centered computing~Empirical studies in HCI}
\ccsdesc[500]{Applied computing~Arts and humanities}
\ccsdesc[300]{Human-centered computing~User studies}

\keywords{Novice Artist, Perception, Generative AI, Interview}

\maketitle

\section{Introduction}

The emergence of Generative Artificial Intelligence (G-AI) has transformed the creative arts landscape \cite{newton2023ai}, impacting literature \cite{li2019storygan,peng2019generate}, music \cite{li2021inco,zhu2022quantized}, painting \cite{brock2018large,karras2019style}, and more. G-AI democratizes the creative process, allowing individuals of any skill level to create diverse forms of art without extensive training. For example, people can transfer famous painting styles to their own art \cite{karras2019style}, generate artistic images from text descriptions \cite{ramesh2022hierarchical}, or co-write fiction with chatbots \cite{brown2020language}. This accessibility opened new avenues for creating and refining art. However, G-AI's rapid advancement has outpaced human capabilities \cite{elkins2020can,haase2023artificial} and raised ethical issues, such as originality, creativity, and copyright concerns \cite{strowel2022study,dien2023generative,liu2023arguments,noci2023merging}. While ethical debates about AI are longstanding, G-AI's unprecedented power has intensified these discussions \cite{roose2022ai,waters2023generative}.

Prior research has explored the ethical boundary from various perspectives \cite{bran2023emerging}. Technically, some suggest reducing the use of human-created content in AI training \cite{abrahamsen2023inventing,bourtoule2021machine}, while others propose practical distinctions between plagiarism and legitimate reproduction \cite{sarkar2023exploring,todorov2019game}. From a social perspective, researchers highlight artist exploitation, not just by G-AI but also by its users and owners \cite{ghosh2022can}. It is evident that G-AI's integration with the art world is not yet harmonious, necessitating further technical, societal, and philosophical efforts \cite{boden2009generative,jordan2020ai}.

Despite the progress, little research considers visual artists' perspectives, which remain crucial \cite{markuckas2022question,bellaiche2023humans,ragot2020ai}. \textbf{In a narrow sense, visual art practitioners refer to those who are engaged in traditional visual art industries such as painting, printmaking, and sculpture. However, in a broad sense, we defined visual art practitioners as those who create visual art works or objects, including automotive design, textile design, ceramics, visual communication, exhibition design, etc.} This definition was similar to the past literature \cite{ko2023large}. Among the artists, \textbf{novice visual artists denote the art practitioners who begin to specialize on their creation process}. They were important to the art society because 1) opposed to established artists, they have the motivation to incorporate AI tools to increase efficiency and create their own art style. 2) they were also more prone to be replaced by AI because of their relatively ordinary creative style. Novice visual artists is a broad group which mainly contained students on their early age of beginning their creative career. To address this gap, our research focuses on how G-AI impacts novice visual artists' careers and their views on G-AI, aiming to bridge these perspectives with future G-AI research. We conducted in-depth interviews with novice visual artists from various specialized fields, leading to these key insights:

\begin{enumerate}
    \item Novice artists have confronted and evolving opinions towards AI-assisted art creation. Besides, artists often hold in-comprehensive mental models towards AI-assisted art creation, which led to disuse of AI tools or trust degradation.
    \item Novice artists integrated AI tools with different manners in different creation stage, which depended on their mental model of adopting AI.
    \item Different stakeholders have different perception of AI usage, which affected art workers' practice.
    \item The future collaboration of novice artists and AI could be divided into two categories: collaboration and having no relation, whereas the public may have unavoidable aesthetic degradation.
\end{enumerate}

\section{Background \& Related Work}

Before the emerging of G-AI, the past literature already investigated the collaboration of humans and G-AIs. Hill et al. \cite{hill2013cost} tested the theory that jointly produced works are of higher quality than individually authored products. Luther et al. \cite{luther2013redistributing} investigated the effect of leadership distribution in online creative collaboration. Cheatle et al. \cite{cheatle2015digital} investigated the collaboration in art furniture production, which integrated digital fabrication tools into studio practice. Later work also explored how to support artists' co-creation \cite{striner2022co} and conducted early stage evaluation. There were also works leveraging digital methods to stimulate the creative process. Ideawall \cite{shi2017ideawall} included combinatorial visual stimuli to facilitate their creative process. 

Regarding the collaboration and evolution of human-AI co-creation, early G-AI facilitates the creation process in the form of GAN \cite{creswell2018generative,goodfellow2020generative} and style transfer \cite{huang2017arbitrary}. Artists, the public and other stakeholders all may use these tools to facilitate creation. Louie et al. \cite{louie2020novice} found the content generated by music AI could make users confused and depressed towards its non-deterministic output. It designed an AI guiding tool to facilitate users' trust towards AI. Menon et al. \cite{menon2024mind} investigated how to include motive sand in ideation and reduce the perception of AI as an un-controlled entity. 

Focusing on the perception of artists, DuetDraw et al. \cite{oh2018lead} investigated the collaboration process between human and AI, which found users were satisfied towards DuetDraw when provided detailed guidance. Besides, they need AI to explain their intention when necessary. The paper presents several implications: 1. guidance and autonomy, 2. dominance of participants, 3. gamefulness through interaction, 4. balance of different functions. Kwon et al. \cite{kwon2023understanding} investigated the inspiration creation process of creators, which found patterns of creators when creating. Han et al. \cite{han2024teams} used prompt co-creation process, finding co-prompting could solve the challenges, but human mostly trusted human partners rather than AI companions. Wang et al. \cite{wang2024exploring} investigated the reflection types and the influence towards users' satisfaction. They found personality is more important than performance. Besides, reflection facilitates programming. Manera et al. \cite{manera2023aesthetic} investigated the aesthetic perspectives of interactive art and text-to-image technologies, which highlighted the exploration of participatory media. 

Specific to some process, Kim et al. \cite{kim2023my} conducted semi-structured interview towards students using AutoDraw, finding AI could raise the perceived security while lacked collaboration, educational support. Kazemitabbar et al. \cite{kazemitabaar2023studying} explored the usage of code generator in programming learning, finding CodeX \cite{finnie2022robots} could help novice programmers in learning. Li et al. \cite{li2024study} investigated students' perception of AIGC assisted design and influential factors. It used smart PLS modeling technique, finding that expectation, social impact and convenience positive influenced the willingness of using AIGC tools. Wan et al. \cite{wan2024felt} found co-writing collaboration could be divided into three process: ideation, illumination and implementation. The future work should enhance the interaction, prompt engineering, transparency and explainability. Ning et al. \cite{ning2024mimosa} as a collaborative music editing tool investigated the human dominated enhancement and creation process, improving the collaborative creation efficiency. Rezwana et al. \cite{rezwana2023designing} introduced a co-creative interaction design framework, which identified pleasure-oriented, improvisational, advisory AI agent. They found most systems lacked effective communication channels. Choi et al. \cite{choi2024creativeconnect} created an AI tool facilitating creation generation, which improved cooperation efficiency through collaborative ideation. 

Regarding the public's perception of the generated content, Rae et al. \cite{rae2024effects} investigated the perception of the public to creators when using AI to assist creation. They hold more positive perspective towards human-created content. The work most similar to us is Shi et al.'s work \cite{shi2023understanding}, which investigates artists' understanding and expectation towards the application of generative AI. They found Generative AI raised the efficiency but also brought competition. Besides, artists and the developers have perception difference towards Generative AI. However, this work 1) did not consider the stakeholders' opinions and their influence on the artists, 2) did not consider the confrontational opinions inside the artists' community. These were clearly addressed in our work. 

% There has been the investigation of designers and artists attitudes \cite{du2023designers} which adopted UTAUT model \cite{marikyan2021unified} or other correlated models.

\section{Positionality Statement}
The primary experimenter is an on-going Ph.D. student in computer science, focusing on human-computer interaction (HCI). He has a foundation in and has conducted research on artistic creation, including work on music generation. During his Ph.D., he primarily worked on HCI and researched the use of large models. He is familiar with the interview group, having had significant interaction with this target group during their undergraduate studies, including activities such as visiting exhibitions together. He has a strong passion for the field of artistic creation and enjoys viewing artworks on display (especially contemporary art pieces). He has engaged in informal studies of painting for a period of time. He understands the confrontations and the refusal reasons of artist. His personal perspective is that there are opportunities for future collaboration between AI and artists. He also hopes that artists can discover potential applications of AI. 

\section{Study Setup}
In this section, we introduced the study setup, which included the experiment design, the participants' recruitment, the data analysis methods and the potential limitations.

\subsection{Experiment Design}
We employed a semi-structured interview format to elicit and capture new perspectives from participants during the experiment, aiming to gather more in-depth information. The interview outline covered six main sections which was inspired by the past literature \cite{shi2023understanding}:

\noindent \textbf{Participant Background:} This section includes the participant's field of study, creative experience, overall attitude towards creation, general attitude towards AI tools, and usage of AI tools.

\noindent \textbf{Participant’s Use and Attitude Towards AI Tools:} This part explores the participant's experience in learning to use AI tools, changes in their views, future attitudes towards AI tools, opinions on whether AI is innovative and capable of creating art, and perceptions of AI's strengths and weaknesses. It also covers the methods, time, willingness, and costs associated with learning AI tools. Additionally, it inquires about the attitudes, experiences, works, and overall views of other creators in the participant’s creative community regarding AI tools.

\noindent \textbf{Participant and Creator Community:} This section examines the attitudes of the entire community towards AI, including creators, clients, and audiences, and how these attitudes are influenced by other stakeholders within the community. It also explores how the participant adjusts their creative strategies under mutual influence.

\noindent \textbf{Necessity, Possibility, and Stages of Integrating AI Tools in the Creative Process:} This part investigates the perceived necessity of AI tools in the participant’s creative process, such as personal creation, client delivery, and completing assignments. If the participant's creative process can be divided into multiple stages, the interviewer will ask about the possibility and reasons for integrating AI tools into each stage.

\noindent \textbf{Participant’s Expectations and Views on AI Tools:} This section asks about the participant's expectations for the future development of AI tools, their attitude towards AI tools, willingness to allow AI tools to use their works for training, desired tasks for AI to replace, and opinions on whether AI is innovative, disruptive, and areas where AI might achieve innovative breakthroughs.

\noindent \textbf{Participant’s Personal Views on AI Tools:} This part uses an open-ended interview format to inquire about any additional perspectives on AI tools that were not covered, aiming to capture the participant’s personalized mental model of AI.

The detailed interview outline is provided in Section~\ref{sec:outline} in the appendix.

\subsection{Participants}\label{sec:participants}
The experiment is grounded in XX university (anonymized for submission), which has the the art college consisting more than 10 departments. Different departments exerted at different perspectives of arts, e.g., Visual Communication, Product Design, Industrial Design, Modeling or Styling, Fashion Design. All participants were beginners in the college (the first year in their undergraduate level), however were skilled in sketching, line drawing fundamentals, and other aspects of painting. 

We recruited 20 participants in total through snowball sampling \cite{goodman1961snowball}. The recruitment process followed data saturation theory \cite{fusch2015we} and grounded theory \cite{walker2006grounded}. When our qualitative analysis reached theoretical data saturation (indicated by no new codes added for three consecutive participants), we stopped recruiting. As our recruiting did not restrict the department and AI usage of participants, we recruited participants with diverse background, AI usage experience and art experience, which enriched our results. Each participant was compensated 90RMB according to the local wage standard. No one dropped the experiment.

\subsection{Procedure}
The experiment was carried out online through Tencent Meeting\footnote{https://meeting.tencent.com/, accessed by Jun 15th, 2024} from May, 2024 to Jun, 2024. For each individual interview, the experimenter first briefed the aim of the experiment to participants. Then the experimenter let the participants signed the informed consent form before proceeding the experiments. Participants were allowed the quit the experiment at any time during the experiment. In fact, no participant quit the experiment. Then the interviewer asks the participant each question systematically, and follows up with additional questions when the participant raises topics of interest to the interviewer. The experiment lasted for 50 minutes on average. All the interviews were recorded and transcribed on a local device.

\subsection{Data Analysis}
We adopted thematic analysis \cite{braun2012thematic} with grounded theory. Two experimenters first jointly coded a subset of the responses (about 20\%, 4 participants) with intermittent discussion to solve potential confrontations. This resulted in the initial codebook. The codes were a combination of structural, magnitude and descriptive codes \cite{wicks2017coding}. Then the experimenters iterated on their codebook through separately coding the rest of the responses given new participants and discussed together. Because we adopted a theoretical grounding research methods, the codebook was consistently updated until the theoretical saturation. At the theoretical saturation point, the two experimenters' inter-rater reliability is 0.90 (calculated using Cohen's Kappa). In the following sections, we provided the themes and sub-themes with detailed explanations and implications.

\subsection{Limitations}

We acknowledge that the research has limitations. First, this study only involved participants from the art college of a single university. Although the participants featured the novice visual artist and the university managed to recruit participants with diverse majors as shown in Section~\ref{sec:participants}, the demographics may not be representative enough for the whole novice artist groups (e.g., some majoring in other art majors, some learning and creating solely by themselves instead of receiving university education). Besides, the potential educational background of the participants may not be the same among all novice visual artists. However, we envisioned our research to serve as the first step towards understanding and facilitating the novice visual artists groups' usage of Generative AI. 

Second, the results may be biased due to recruiting and social desirability \cite{grimm2010social}. Those who reluctant to share their viewpoints about AI and those who holds negative opinions towards AI tools were less likely to participate in the experiments. We tried our best to understand these minor attitudes through asking each participant about their community's opinions such as other students' negative opinions. However, this only resulted in a third-perspective viewpoint. Besides, participants may tend to disclose more positive opinions during the interview due to social desirability effect \cite{grimm2010social}. Thus, the objective attitudes of participants may be more pessimistic.

Third, although we tried our best to diversify the departments in order to include all fine art majors, the results may be specific to a subset of fine art majors, e.g., easel painting. Different art majors may have different creation ways. However, as we included different majors, other majors may involve more or less AI with similar manners. We expected our results could inspire later works for designing better collaboration tools and interactions between AI and novice visual artists.

\section{RQ1: Evolution of the Opinions towards AI Tools}
Artists have two main objectives to create art works: to create for themselves and to create for the work. The former aim usually would not incorporate AI tools. This is because artists have their own emotion and feelings to express. They think AI could not replace their want of drawing by hands, which is echoed in the past literature \cite{notaro2020state}. However, for the objective of creating for the work, artists usually aim for higher efficiency, create innovative and influential works, etc. Thus, there is potential space for AI tools \cite{shi2023understanding}. In the rest of the paper the analysis would mainly focus on creating for work without specification. 

\subsection{Evolution}

\begin{figure}  
    \includegraphics[width=0.6\textwidth]{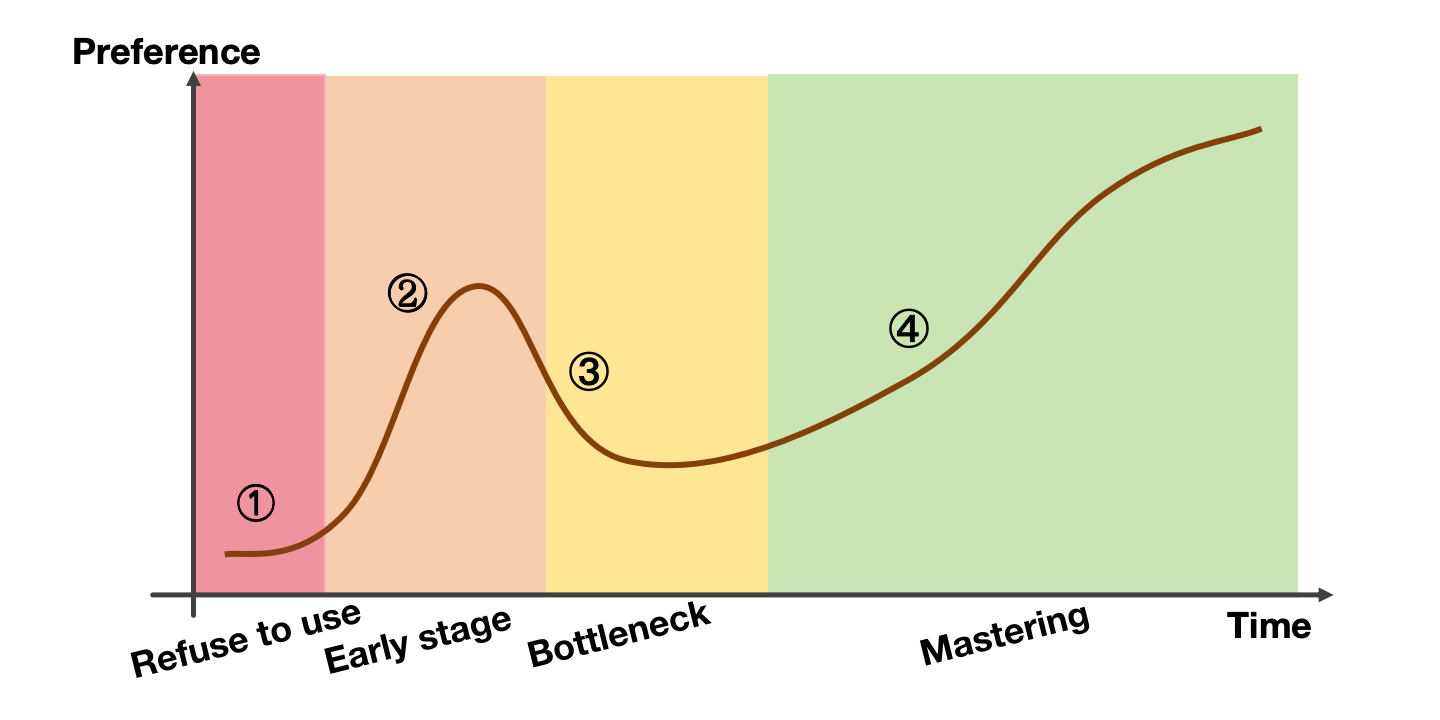}
    \caption{The different stages of art practitioners' practice of AI-assisted creation. 1) Before usage, art practitioners were negative towards AI's usage due to different pre-conceptions. 2) Upon usage, they found AI could complete some basic tasks for them and thus increased their preference. 3) However, with time, they found the tasks that AI could not complete could not easily be solved. Besides, the capability of AI still could be satisfy their need and aesthetic requirement at most times. Thus they decreased the preference. 4) Some practitioners keep on interacting with AI tools and gradually mastered the generation skills. Their preference increased steadily before converging with the increasing of art work's quality.}
\end{figure}

Similar to the technology acceptance model \cite{moore1999crossing}, the change in perspectives on AI tools can be divided into three stages over time. The first stage is when there is no usage at all, the second stage is when there is some usage and experimentation, and the third stage is when the usage is proficient. In the first stage, preconceived stereotypes about the technology often influence users' views and opinions on AI tools. Because most initial artists hold negative stereotypes about AI tools, they typically believe that AI tools should not be used. They consider AI-created works to have copyright issues, low quality, and unable to truly become art. 

The second stage is when participants began to use the AI-assisted tools for creation. In this stage, most participants foresee the capability of AI to generate well-formed drawings efficiently. All participants without except were astonished at the speed of the generation. \textit{``It would be exciting if they could be used for improving my workflow.'' (P2)} However, they were usually frustrated during their trials. \textit{``I could never let the AI understand what I want to draw, although I have seen aesthetically appealing drawings of AI.'' (P3)} 

The third stage is when usage becomes proficient. By this stage, artists have usually accumulated significant experience through extensive and long-term use of AI tools, allowing them to effectively leverage the strengths and capabilities of AI tools. The transition from the second to the third stage may simply be due to the accumulation of time, or it could be due to opportunities for large-scale use of AI tools (e.g., internships, projects, etc.). In this stage, users' attitudes towards AI tools are more positive in some or all aspects, accompanied by a frequent usage. They believe that AI tools can achieve higher efficiency than humans in certain areas, and in some cases, even produce higher quality works than human creations. \textit{``AI could help me construct the 3D models with very high efficiency. The drawing quality is also higher than me.'' (P2)} However, some artists maintain a relatively conservative attitude, believing that AI creative tools are still limited in many aspects, such as physical sculpture or weaving.

This conceptual shift and reconciliation, as well as the neutral or impartial attitude towards AI, may not necessarily be achieved through further in-depth use of AI. In many cases, users do not engage in deeper or significantly more extensive use of AI, nor do they have a clear understanding of its broader applications. However, their perceptions have still softened to some extent. Perhaps they recognize the irreversible trend of AI's significant development, understanding that the resistance from the artist community is relatively ineffective compared to the broader support from the general public. The wider population finds many conveniences in using AI. Even those unfamiliar with AI technology can derive a sense of achievement and participation from its use. Many people are employing AI for various tasks, such as writing drafts, composing reading reflections, and even creating music. These experiences lead them to acknowledge the objective existence and progress of AI. Consequently, their previous anxiety, unease, and rejection gradually diminish as they continue to reflect on the matter. Although many artists still resist and emotionally dislike AI, this resistance might stem from their beliefs and stubbornness.

\subsection{Refuse Reasons \& Disadvantages}

% 很难保证角色一致性，如果是一系列人物就很难固定。
% 画错很多东西，每个地方都画很多东西，头发划分成两大块，没有清晰的逻辑。
% 二分阶段，只停留在二分阶段，如果成图就不是很好改。

The reasons for the rejection of AI by artists can primarily be attributed to the following:

\noindent \textbf{Artists believe that AI-generated works involve copyright issues}. Specifically, many AI models are trained on human artists' works without permission (at least, this is their perception), making AI-generated works a form of theft of the artists' labor and intellectual property. This was also contextualized and discussed as value misalignment in the past work \cite{allred2023art}. 

\noindent \textbf{Artists argue that AI-generated works lack originality}. Because the training data consists of human artists' works, AI creations are seen as a patchwork, a collage of human-made elements devoid of soul, style, creative intention, and thoughtful artistic process. This is also reflected that AI would always generate pictures within a small subsets with similar styles. For example, AI would always generate images that appear younger, sexualized and lighter skin tones \cite{lovato2024foregrounding,luccioni2021s,ghosh2023person}. \textit{``The AI has their own fixed style which is greasy and lacked variation.'' (P8)}

\noindent \textbf{Artists contend that AI-generated works often fail to meet their expectations in terms of quality and aesthetics}. For instance, artists typically require effective handling of details, light, and shadow in their works, along with a thoughtfully designed composition. AI-generated images, however, are often only high in completeness but lack meticulous refinement, making them difficult to use directly. \textit{``AI highlights every part of each picture and renders every detail meticulously, which is not what drawing should be about.'' (P8)} Besides, they usually could not easily modify the drawings because AI generated the picture in a whole instead of in a layer-wise manner. \textit{``It directly output the whole picture and even if with photoshop modifying the original picture is hard.'' (P16)} They could only discard unsatisfying drawings. This further intensified the problems that selecting qualified drawings from numerous unsatisfying drawings is time-consuming and tedious, which hindered artists' usage of AI.

\noindent \textbf{Artists feel that using G-AI for creation is too easy}. G-AI can quickly complete a piece of art, significantly reducing the labor time of human artists, which might be seen as a form of laziness. Most drawings with G-AI were recognized as without artists own thought. The paintings were deemed as rushed due to a lack of time. Artists believe that using AI in their work does not embody their own effort and ideas. Most G-AI did not have a physical body, and nor could they perceive the physical world. \textit{``Artists used G-AI because they have no time to think, perceived and draw themselves.'' (P8)} Even if some specific AI could perceive the physical world, they could not generate. This was also linked to some common pre-conceptions that artists or people would naturally linked AI-generated art to inferior art \cite{hong2019artificial}.

\noindent \textbf{Artists think G-AI's processing logic is different from theirs.} Artists thought G-AI complete a drawing with all details in a short time, which is different from humans. The G-AI tools made them unable to understand AI drawings and modify the generations. Besides, artists interleaved drawing and thinking, which made the direct completion of G-AI infeasible. \textit{``I only think out a draft before starting to draw. I refine my thoughts upon the drawing process.'' (P18)} Despite all these, the G-AI's drawing is too detailed to be aesthetic appealing. The drawings of human visual artists leave blank spaces and highlights the important element in the drawing. These enhanced the balance and harmony of the painting.

% 让艺术的门槛降低了，但是国内的美育工作做得不好。因为地区和年龄因素导致审美和美育工作做的不到位。AI过于开放会导致AI绘画门槛降的过低。深化美育差距，旱的旱死涝的涝死，审美降级。
% 没有艺术表达，没有灵魂。

% AI画的图没有涂层的概念，和背景很难分开。

\subsection{Accept Reasons \& Advantages}

Notably, artists who hold a positive attitude towards AI mostly turn their attitude from negative to positive through the usage. They admit they hold similar refusing beliefs and dispel the aforementioned pre-conceptions upon usage. We detail the reasons, which is also the advantages as follows:

\noindent \textbf{AI could produce remarkably stunning works}. While most artists still recognize that AI lacks originality \cite{sarkar2023exploring}, they also admit that AI can sometimes produce remarkably stunning works. This stunning quality is reflected in the composition, coherence, layout, and aesthetics of the artwork. \textit{``Sometimes AI really makes me inspiring through combining different chunks in that unexpected way.'' (P19)} Although the result may encapsulate human artistic endeavors, it also introduces some novel elements. However, artists remain relatively conservative regarding the copyright issues, quality, and aesthetic standards of AI-generated works. Most artists believe that AI-generated works potentially contain copyright issues and that their quality and aesthetics may not necessarily meet human expectations and standards.

\noindent \textbf{AI could complete the picture efficiently, which often surpassed human's speed with a comparable quality}. When asked about the aesthetic value of the paintings, artists reveal that \textit{``although AI paintings is not supreme in quality, they could still complete the picture with high efficiency.'' (P20)} AI tools can produce well completed works in a short time, including but not limited to ordinary images and 3D models, with significant detail, lighting, and color coordination. This facilitates production, especially when artists need to create complicated work in a short time. This was also echoed in the past work concerning design processes \cite{pearson2023rise}.

\noindent \textbf{Using AI for creation is not easy}. Artists acknowledge that creating works with AI is not easy; at the very least, producing work that satisfies them is not a trivial task. It requires learning some knowledge, experimenting with adjusting parameters, and so on. This finding contradicts the common perception of artists that incorporation of AI is a lazy behavior because it quickly complete a painting.

\noindent \textbf{AI act as a source of inspiration}. Some artists find that AI-generated works can provide inspiration and creative ideas. Especially when they used AI to generate artworks in an early stage of creation, they often could get un-anticipated inspirations from AI-generated work. This is usually through ``generating numerous candidate pictures and select the ones with higher quality.'' (P9)

\noindent \textbf{The AI artwork exhibited overall cohesion with high quality}. A few artists believe that AI-generated works exhibit excellent overall cohesion and structure, often surpassing human-created works in terms of layout and composition. These could serve as references for practitioners' creation. \textit{``I think AI cope with the layout and the understanding towards the spatial relationship with high quality. I could also learn from its thoughts.'' (P6)}

Although the copyright issues are hard to solve, artists still show tolerant attitudes. They provided their attitudes which is summarized in Section~\ref{sec:discussion}.

\subsection{Mental Model}

Owing to the opaqueness of AI \cite{bryan2023explainable}, the mental models of AI' generation process was various and innovative \cite{dipaola2023art}. Regarding the generation of AI, there were 2 mental models of participants, whereas regarding the training process of AI, there were 3 mental models.

Regarding the generation of AI, some participants (16/20) thought the artwork was directly generated without the iterative denoising process. This made them astonishing at the efficiency of AI and uncomfortable towards the generation manner of AI, which allowed no modification space. Other participants (4/20) thought the artwork was generated through iteratively creating and modifying the artwork on some random points. With numerous steps' of iteratively modifying, the noise would finally turn into the artwork. 

Regarding the training process of AI, some participants (8/20) believed AI simply copy patches of the training dataset to the artwork to create a new work. This is also denoted as ``stitched corpses.'' Owing to this mental model, these participants were strongly against AI-generated work because it mechanically and inhumanely pieces together works from different artists, lacking morality. They contend that \textit{``such works are disjointed, with parts resembling one and other parts another, resulting in a creation devoid of a unique artistic style.'' (P12)} Also some participants (12/20) believed that the AI learned old patterns and generate new artwork patterns through mimicking the old patterns. As P5 said, \textit{``the AI extracted the patterns and correlations of different pixels and instantiated them on a new picture.''}, these participants thought AI learn abstract patterns (e.g., pixel patterns, stroke patterns) from the training dataset and generates new artworks based on patterns. A few participants also mentioned (4/20) that AI and human creates artworks in a similar way. Both AI and humans go through a process of gradual refinement. AI can be improved step by step through prompts (image and text) provided by humans. Similarly, when humans paint, they start with a concept or a rough draft and then iterate, refine and perfect their work based on ideas that arise during the actual painting process and detailed needs. As AI's learning process leveraged a larger dataset, it potentially could create better artworks than human practitioners.

\subsection{Discussions}\label{sec:discussion}
There were wide discussions about the perception of AI. This study provides the first-step inspiration through selecting the two most mentioned aspects \cite{shi2023understanding,nguyen2023effects}.

\subsubsection{Copyright Issues of AI-generated Work}

The copyright issues were widely discussed in CSCW and art community \cite{nguyen2023effects} and intensified with the emergence of AI techniques \cite{fiesler2016reality}. We identified the reasons and opinions towards copyright issues.

\noindent \textbf{AI used unauthorized artworks for training.} 20/20 participants agreed that AI would use un-authorized artworks for training. They gave several examples where they heard of or experienced that AI could generate the AI style similar to famous artists. This exemplified their pre-conception. \textit{``I often see the famous artists' paintings appeared in the output of the AI. I even see many LoRas with the painting style of just that painter. But I know that specific painter did not provide the copyright.'' (P13)}

\noindent \textbf{AI simply copys what they learn without modifying.} 8/20 participants also agreed that AI differs from human in how they copy with copyright issues. The learning process of AI is akin to that of human, which both involved imitation of exemplary works. However, when novice visual artists create new artworks, they were taught to create distinct artworks rather than replicating identical ones. However, AI merely imitates and learns from these works without innovation. Consequently, even if artists' style may overlap, their works still retain their own unique thoughts, whereas AI were criticized of merely plagiarizing human creations.

% Many artists believe that AI has copyright issues, a view similar to their perception of AI as assembling pieces from various sources. In reality, the learning process of AI is akin to the human process of learning and imitation, both involving the selection and study of numerous exemplary works. The primary difference between AI and humans lies in their approach post-learning: humans strive to develop their own creative style after studying selected works, while AI merely imitates and learns from these works without fundamental innovation. Consequently, this is why many artists' styles may overlap but still retain their own unique thoughts, whereas many subjects believe that AI lacks originality and is merely plagiarizing human creations. The copyright issues have been discussed in CSCW and art community \cite{nguyen2023effects}, however were intensified with the emergence of AI techniques \cite{fiesler2016reality}. 

\subsubsection{Whether AI has Creativity}

It is debated whether AI has creativity \cite{seo2021study,hwang2022too,xu2020discussion,browne2022or,hong2019artificial}. From the interview, there were two themes AI has no creativity and three themes debating AI has creativity. 

\noindent \textbf{Pro: AI provides inspiration through assembling works.} 5/20 believe although AI has not yet created accepted art styles, it integrates, collages and trims existing styles. The produced work often provide artists with new inspiration in composition, color schemes and creation. This inspiration is similar to those provided by innovative works.

\noindent \textbf{Pro: Art is revolutionized by technological advancements, thus the advancement of AI fertilizes artworks.} 3/20 believe the progress of artistic styles such as hand drawing, sketching, Impressionism, and Post-Impressionism is driven by technological advancements. These styles emerged due to human's deeper understanding of light, imaging rules, and lighting effects, enabling better depiction or construction of various light effects. This understanding of real-world imaging also provided artists with new ideas, driving the transformation of art forms. AI, with its superior capabilities in technology and logical reasoning, might discover new art forms based on new technologies and a new understanding of the world \cite{guan2023leveraging}.

\noindent \textbf{Pro: Art processes and produces work in a manner similar to human, thus enables creativity.} 4/20 participants argue that the process of AI learning to paint is similar to that of humans. Both involve learning patterns and seeking regularities from data. From this perspective, AI's learning process can incorporate a larger dataset, potentially achieving better results than human learning. AI also has a greater likelihood of reaching or even surpassing human levels of creative output. \textit{``AI would extract the pixel patterns and learn the abstract representations from the past work. I think I understand right from the technical perspective. Thus from the technical perspective you could expect AI has creativity.'' (P10)}

\noindent \textbf{Con: AI rearranges existing art styles and assemble artworks.} 6/20 art practitioners think that AI could only learn from the data and mechanically assemble different patches from different artworks into a new drawing. It is far from innovation. \textit{``You could not say it is real creativity because it is only copy pasting. If you say copy pasting is also a creative style, then AI even have no idea or conscientiousness of this copy pasting.'' (P14)}

\noindent \textbf{Con: AI did not think and express proactively.} 4/20 art practitioners think that AI could not think and express their emotions like human, which limited their innovative thoughts. This echoed the previous work \cite{xu2020discussion}. Artists thought it is attributed to current AI's inability to engage in critical thinking. For instance, P5 mentioned Andy Warhol's portrayal of Scarlett. She thought it is a re-creation of traditional artistic imagery using inexpensive paints. That subverted past aesthetics regarding the image and the use of materials, thus constituting innovation. However, AI cannot break such inherent conventions and essentially engages in repetitive learning. 

Different from the reasons mentioned in the past work \cite{kahney1986problem,SIMON1973181}, we identified novel reasons from the perspective of novice visual artists perspective. There were positive opinions towards the creativity of artworks. However, whether audience acknowledged the price of AI's creativity \cite{guerreiro2022using} and what effect it would have on artists' creativity \cite{horton2023bias} remains un-examined.

\section{RQ2: Practices of AI Tools}

\begin{figure}[!htbp]
    \includegraphics[width=0.9\textwidth]{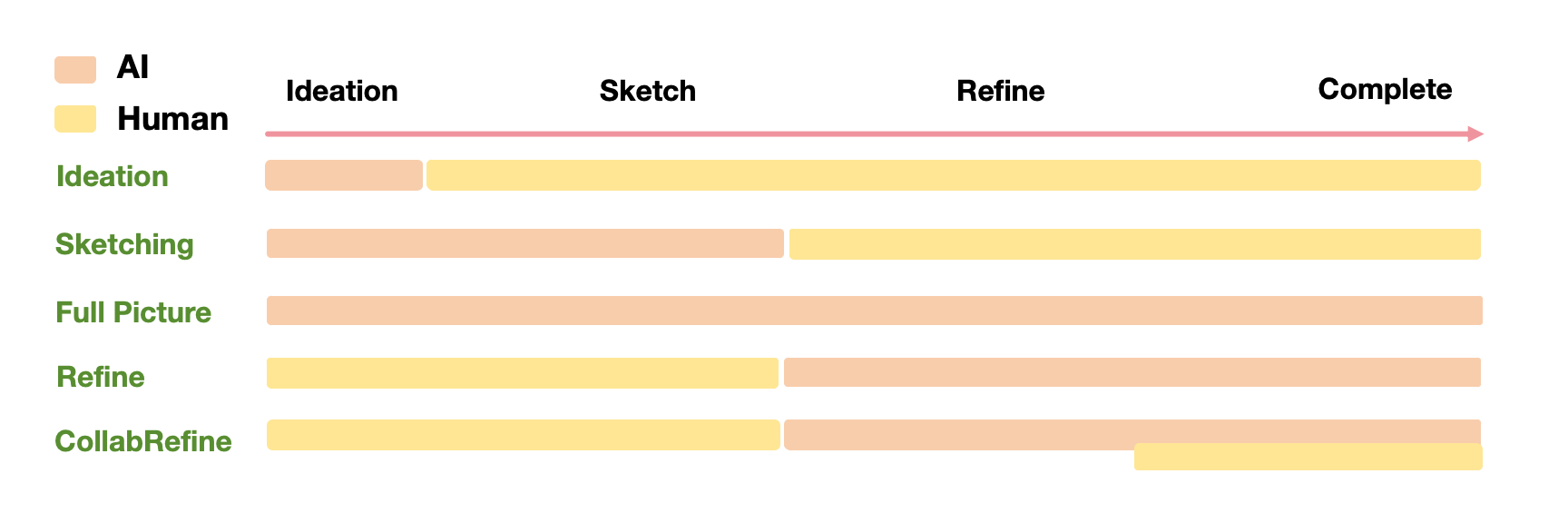}
    \caption{The different collaboration modes where AI participate.}
    \label{fig:ai_participation}
\end{figure}

The practices of AI tools were analyzed along the artists' creative process. Different majors may have different creative process. However, the process could be generally divided into the stages in Figure~\ref{fig:ai_participation}, which was similar to the past literature \cite{botella2013artists,botella2018stages}. 
% the past literature divides the creative activities into several stages: general ideas or vision, documentation and reflection, first sketches, testing forms and ideas, provisional object and draft, final work and series.
\subsection{Cases Needing AI?}

% AI建模可能帮助，省力气，苦力，可以用AI代替

The demand for AIGC tools varies with different tasks. For work-related tasks, artists are more inclined to use AIGC tools. In these tasks, their role is primarily to create content that meets specific requirements and guidelines, with less emphasis on self-expression. In such cases, while some tasks still require manual completion by artists due to AIGC's inability to generate satisfactory images. However, certain tasks, such as in betweening or non-drawing-related content, are also deemed feasible for AIGC to automatically complete. \textit{``I think for those ideation work, AI could complete far better than human, at least for work.'' (P7)} This trend might become more prevalent in the future. Additionally, the limited use of AIGC tools in work-related tasks is not due to their ineffectiveness but because their usage is easily detectable and may lead to ridicule from peers or be prohibited by companies. These social factors hinder the widespread adoption of AIGC in the field of artistic work.

Novice artists majoring visual communication are likely to be more interested in AI, especially in the contexts of painting and poster design. These areas should be distinguished, as fields such as textile design might not see as significant a difference, and thus the impact is less pronounced. The need for AI varies depending on the specialization and individual preferences. This greatly extended the past work that only mentioned AI were important in classes' education \cite{chiu2024artificial}. Whereas for novice visual artists specializing in ceramics, many believe that despite the need for a physical entity, there are numerous situations where their creative process might involve the use of AI. Particularly during the 3D modeling phase, AI can provide them with a wealth of 3D model inspirations, offering more space for contemplation. \textit{``I think maybe AI could at least help me construct or sketch the outline of what I need to complete. Then I could sculpt myself.'' (P16)} While this may not be directly related to the actual creation process, it remains significantly important.

\begin{table}[h!]
    \centering
    \caption{The potential cases needing AI according to the workflow of different occupations (art majors). The results aggregated the utterances of different participants.}
    \label{tab:sample_table}
    \begin{tabular}{l l}
        \hline
        Art Major & Needs \\
        \hline
        Information Arts and Design & whole workflow \\
        Fashion Design & ideation \\
        Exhibition Design & ideation, graphic design \\
        Product Design & background investigation, ideation, graphic design, 3D modeling \\
        Industrial Art & ideation, graphic design, 3D modeling \\
        Visual Communication & ideation, graphic design \\
        Sculpture & none \\
        Fine Arts (Painting) & ideation, main process \\
        Easel Painting & none \\
        Experimental and Fine Arts Major & background investigation, ideation \\
        Mural & ideation \\
        Ceramics & 3D modeling \\
        Automativ design & ideation \\
        Dyeing and weaving & none \\
        \hline
    \end{tabular}
\end{table}

\subsection{Cases and Reasons without AI?}

Most artists believe that incorporating AI into their creative process is unnecessary. This is because artists have their own thoughts and ideas even for their work. These thoughts and ideas are difficult to convey to AI and cannot be easily translated into language. Moreover, AI struggles to comprehend the emotions and ideas that artists wish to express. \textit{``I think drawing is not something which could easily be expressed using words. Otherwise it is somewhat horrible.'' (P11)}

Additionally, the creative process for many artists is akin to the basic needs of eating and drinking. Even for work, their creating could not be distinguished from their own life and emotion. \textit{``Just as hunger prompts eating and thirst prompts drinking, the urge to create prompts artistic expression.'' (P1)} This intrinsic need cannot be met or replaced by AI, nor can AI even comprehend it. From another perspective, AI cannot generate these emotions or needs, nor can it use such emotions and needs to support its own creative process. Consequently, most artists believe that AI cannot participate in or replace the self-expressive aspects of artistic creation in their work. AI is \textit{``more suited as a tool in the production or purpose-driven aspects of the creative process to increase efficiency'' (P11).}

\subsection{When and How AI Participate}

Based on the varying degrees of involvement in the creative process, AI-assisted creation can be categorized into three levels: AI as a conceptual aid, AI as the main creative foundation, and AI as a full or partial replacement of the creative process, which was also reflected in Figure~\ref{fig:ai_participation}.

In the first level, the artist initially sketches or drafts a few lines, then uses AI tools (such as Stable Diffusion, ControlNet, or other tools) to generate multiple works. The artist draws inspiration from the generated works in terms of color schemes, composition, and other elements, continuing to create based on their initial sketches or drafts. Here, the AI-generated works do not appear in the final creation but serve solely as sources of inspiration. However, the artist may mimic or learn from the AI's style, resulting in a final piece that resembles the AI's suggestions.

In the second level, AI serves as the main creative foundation. The artist starts with a basic draft or sketch and then lets the AI generate multiple works. The artist selects one of these AI-generated works and further refines it by adding layers or making adjustments. This process enhances the stylistic diversity, quality, and overall aesthetic of the work. The final product is a collaborative piece refined by both the artist and the AI. This is also due to the fact that AI can currently only complete some preliminary tasks in certain artistic fields. For example, in the sculpture industry, AI can only assist with the initial conceptualization work and is not capable of performing the actual carving.

In the third level, AI functions as the primary creative tool. The artist provides a basic draft, and the AI completes the remaining steps. The artist then submits or presents the AI-generated work as the final product. \textit{``I sometimes directly hand in AI-generated work if there is no time to modify. I hope the receives did not find I use AI.'' (P13)}

AI can also assume various roles in the artistic creation process. Some AIs function as tools, primarily aiding in tasks such as coloring and expediting the painting process. Others act as mentors, providing guidance on composition, inspiration, and other elements, thereby helping artists to perfect their work.

\subsection{Why Different Collaboration?}

The primary determinants of different collaboration models are as follows: 

\noindent \textbf{AI capability and perception of AI capability}. Some artists' resources and creative abilities do not allow for AI-generated high-resolution works. \textit{``I do not have graphical cards supporting generating drawing larger than 1024x1024, which make the AI drawings hard to be applied. On canvas, on the opposite, I can draw arbitrarily large.'' (P5)} Enlarging works of ordinary resolution on canvas often results in poor appearance and blurred details. Therefore, these artists opt for hand-drawn sketches, using AI to provide creative inspiration. Additionally, some artists believe that AI-generated works have many flaws and errors that necessitate final corrections by hand, leading to a workflow where AI generates the initial version and humans refine it.

\noindent \textbf{Copyright issues}. Some artists are aware of potential copyright problems with AI-generated works, thus they do not use AI for direct creation. Instead, they primarily rely on their own brushes for creation, allowing AI to provide only inspiration; alternatively, artists might manually make extensive revisions to AI-generated works. 

\noindent \textbf{Desire to conceal AI usage}. Due to negative perceptions from the artist and audience communities, some artists prefer that their works do not reveal AI involvement. Therefore, many artists use extensive revisions and complete the later stages of the work themselves after initial AI inspiration as a means of self-protection. Only through such a creation process can AI-assisted works be suitable for public release. \textit{``I don't want others to find I used AI. Thus editing at the last step is necessary.'' (P7)}

\section{RQ3: The Stakeholder's Opinions towards AI Tools}

Artists shared their perspectives on public opinions regarding AI tools, identifying three primary groups: clients (those who commission artworks), the audience, and the creator community. Compared with the previous literature which only covered the artists' own perspective or the public's perspective \cite{bellaiche2023humans}, we explicitly proposed the triad relationship among different stakeholders towards AI-assisted visual art creation.

\begin{figure}[!htbp]
    \includegraphics[width=0.7\textwidth]{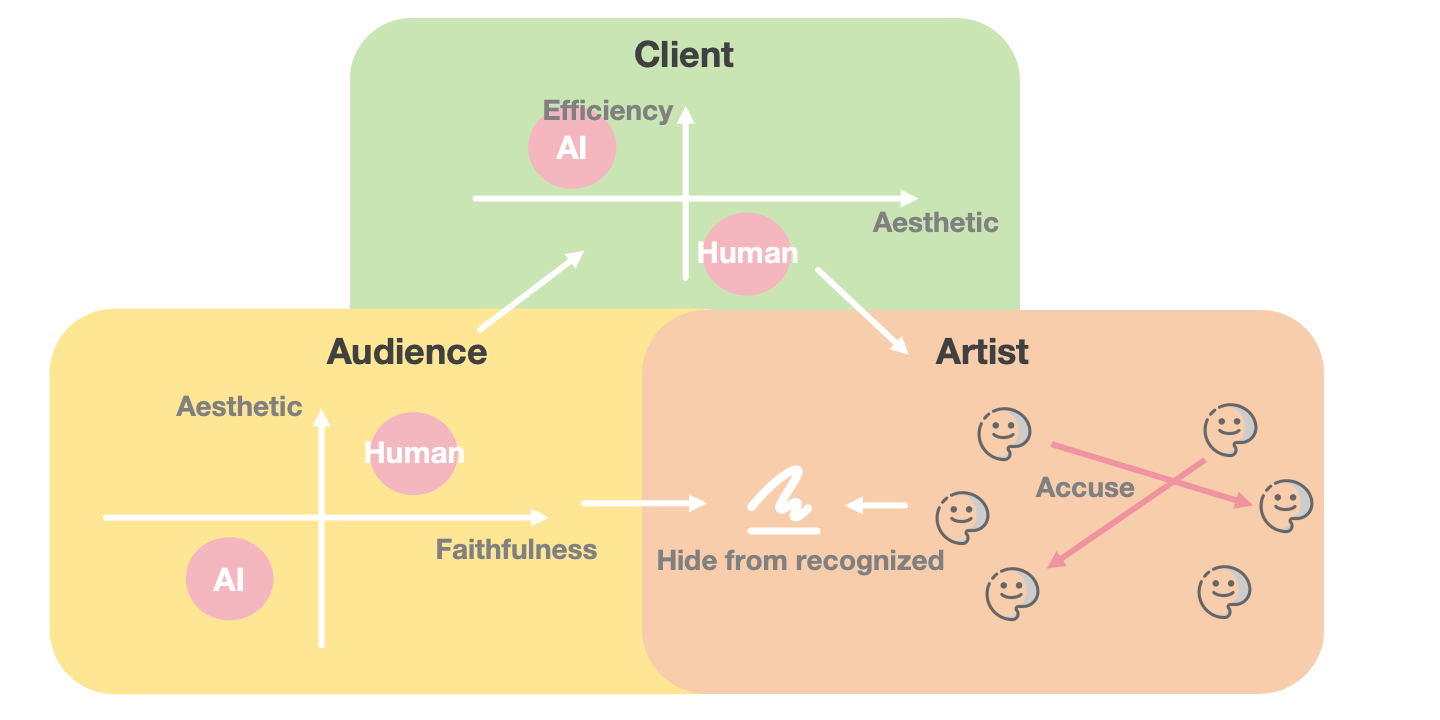}
    \caption{The viewpoints of different stakeholders: clients, audience and artists.}
\end{figure}

\subsection{Clients}

Clients include two types: individual clients and corporation clients. The individual clients primarily include other artists, fans of the specific artists who commission the artworks on platforms. The corporation clients include government agencies, profit-oriented companies, individual entrepreneurs, etc. 

Clients' view was mainly influenced by their 1) pursue of efficiency, 2) own aesthetic value. Technologically enthusiastic clients may permit or even encourage artists to use AI, appreciating the efficiency and time-saving benefits. \textit{``The previous client encourage me to use AI and even teach me how to better incorporate AI tools into the workflow when I create the LOGO for them.'' (P9)} AI can also facilitate rapid accumulation of creative ideas. Conversely, more conservative clients, especially those with lenient requirements and lower aesthetic expectations, may allow AI usage due to the difficulty in discerning AI-generated content. Clients with stringent quality demands prefer artists not to use AI, or at least ensure its usage is undetectable. These client perspectives generally align with audience and societal opinions. While many clients favor AI for cost reduction and efficiency, numerous audience members oppose the integration of AI traces in artworks. Additionally, some art critics and fraud exposers actively identify and disclose AI usage, potentially leading to public backlash, thus hindering creators' use of AI tools. 

\subsection{Audience}

The audience for artworks comprises those who receive and view them, and their perceptions vary based on two primary factors: the platform and their personal preferences. Previous studies have demonstrated that AI influences the evaluation and reception of AI-generated artworks from a general perspective \cite{gu2022made}. On personal commission platforms, most audience members prefer handmade creations and specific artists' styles, showing low tolerance for AI involvement. Fans of particular comics or artists also share this view, rejecting any AI traces in artworks. \textit{``I also commission works from my favorite artists on the platform and play RPG games daily. I cannot admit the existence of AI assited painting because their were too aesthetically unappealing.'' (P10)} Conversely, the broader audience often remains neutral towards AI usage, generally unable to distinguish AI-created elements from human-created ones.

The audience's perspective on AI art is significantly influenced by the genre of art they appreciate. For instance, the ``Anime, Comic and Games'' (ACG) community tends to have a predominantly negative view of AI art, preferring works created by human artists. This preference extends to other works incorporating ACG elements, such as movies, anime, and certain video games. For games targeting an ACG audience, the use of AI to generate primary character designs or artistic effects can significantly impact user opinions and perceptions. Conversely, audiences in other fields exhibit relative neutrality, displaying neither strong preference for nor aversion to AI-generated art.

From the artists' perspectives, many believe that most people cannot discern between human-created and AI-created art, a view supported by past literature \cite{chamberlain2018putting,gangadharbatla2022role,mazzone2019art}. Regarding aesthetic values, some studies found human-created art inferior to AI-created art, while others did not \cite{hong2019artificial,israfilzade2020s,xu2020using}. Additionally, the perceived increase in the value of AI was not echoed by artists, possibly due to the lack of perceived autonomy and competence in AI-generated works \cite{latikka2023ai}. Our findings indicate that audience perception correlates with audience types. Although distinguishing between AI and human artworks may currently be challenging for the general audience but easier for artists, this differentiation may become more difficult in the future due to advancements in AI technology \cite{oksanen2023artificial}.

Furthermore, the public's opinion towards generative AI-assisted creation differs from their opinion of generative art. Generative art, created by established artists using AI as a tool, incorporates their personal thoughts and creativity \cite{alves2023generative}.

\subsection{Creator Community}

The creator community generally holds a resistant and negative stance towards AI. Many artists do not appreciate AI-generated works, the reasons of which were similar to their own refusal reasons. Some view AI-created works as a form of laziness, reducing the time and effort traditionally invested by artists. Others criticize AI-generated works as cheap and crude despite their intricate details, noting incorrect lighting effects and a perceived ``cheapness'' associated with AI. \textit{``These clients usually give lower price for AI-assisted works because they AI could complete the work faster. We also think these works were cheaper.'' (P20)} Additionally, concerns about potential copyright infringement contribute to the community's reluctance to embrace AI. During interviews, even in peer discussions, AI-created examples were notably absent. Most artists engage in personal AI experimentation but refrain from sharing these works publicly or discussing them with peers, seldom receiving positive or constructive feedback. However, they were active in criticizing the AI-assisted artworks which were of low aesthetic value, as most often regard themselves as with obligations. 

% 感觉还是抢不抢饭碗的问题

\section{RQ4: Expectation and Confrontation towards the Future}

One notable characteristic of novice artists is that while they possess strong painting skills, some were not familiar with future career paths. This resulted in two consequences:

\noindent \textbf{On one hand, this might lead to a superficial or erroneous understanding of how AI could replace or integrate into future creative processes.} Some artists, unfamiliar with specific steps in the creative process, may view the integration of AI into their future work negatively, even believing that AI cannot be incorporated into their creative workflows. \textit{``I do not know much about what in the future I will do, so it is hard for me to say the relationship with AI.'' (P13)}

\noindent \textbf{On the other hand, they are less constrained by existing workflows and might bring more innovative ideas.} For example, one artist mentioned in an interview the possibility of generating 3D models directly from text, thereby replacing traditional scene-building processes. He admitted to not being very familiar with the conventional scene-building process, but emphasized that using text to generate 3D models could achieve a high level of completion in a short time and produce multiple samples, significantly enhancing productivity. \textit{``I think the future workflow could be replaced partially by AI techniques, even for those 3D modeling tasks. AI definitely will evolve in the future.'' (P9)}

With AI-generated content (AIGC) showing up in the art market and gaining increasing recognition and acceptance \cite{haase2023artificial}, some believe that jobs involving a substantial element of design or writing are at risk \cite{waters2023generative,zarifhonarvar2024economics}. While negative voices towards G-AI taking away the artists' market are heard, some researchers \cite{epstein2023art,zarifhonarvar2024economics} see the threat of job replacement in short terms but the enabling new forms of creative labor and reconfiguration of the art economy as the ultimate result.

\subsection{Confrontation or Tools?}

\begin{figure}
    \centering
    \includegraphics[width=0.9\textwidth]{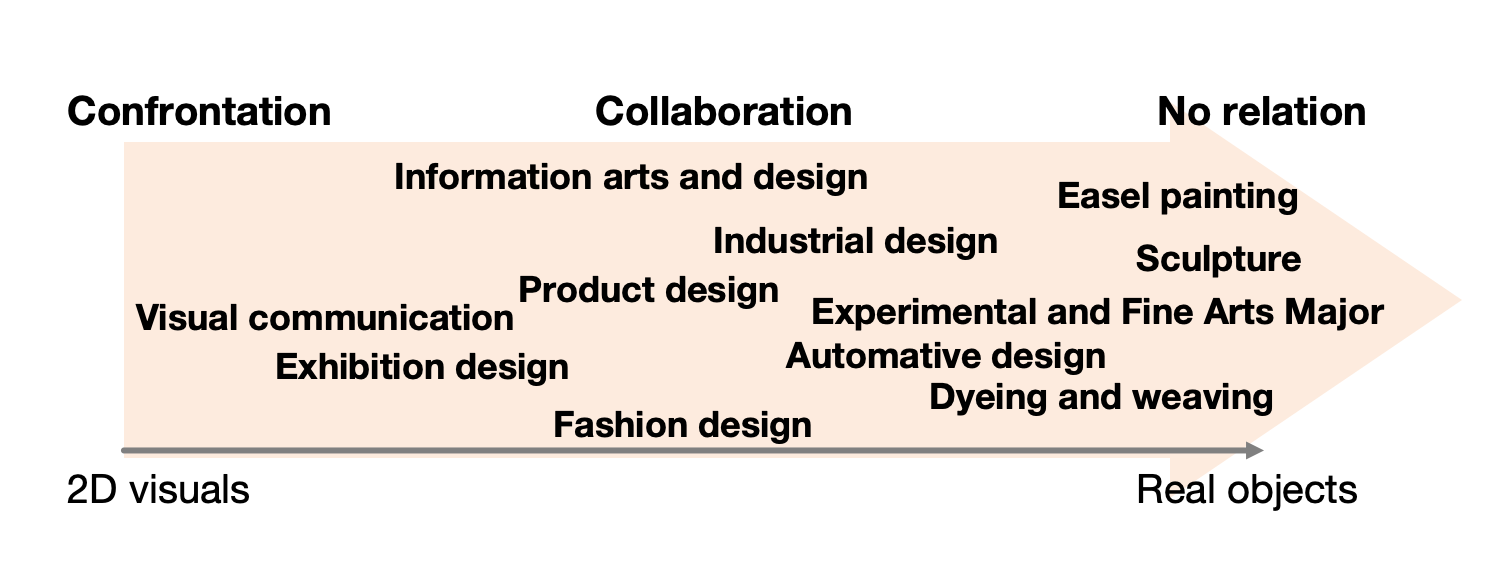}
    \caption{The potential confrontation and collaboration of different majors, represented by spectrum graph.}
    \label{fig:art-collaboration}
\end{figure}

There is some fear about whether AI would finally replace human in the arts fields \cite{hong2019artificial,tubadji2021cultural}. Regarding the future capabilities and potential of AI, creators most frequently discuss the relationship between AI as a tool and themselves. Almost without exception, artists hope that AI can serve as a tool to assist them in completing certain tasks and concepts, but not as a replacement for humans, especially not in competition with their manual creation of works. Astonishingly, those majoring in product design or information arts thought \textit{``Easel painting were the major mostly likely to be replaced because their creation form were more similar to the paintings of AI.'' (P2)} However, the participants from painting major and fine art major did not think so. \textit{``Painting is something which only artists could do because it involves the perception of the objects. These design-related majors were more likely to be replaced.'' (P11)}

Nevertheless, due to the rapid development of AI, many artists are reluctant to feed their works into AI training systems, as they believe this would enable AI to effectively mimic their artistic style, producing similar creations and thus becoming a competitor. \textit{``I definitely do not want my works to be replaced.'' (P8)}

Artists believe that AI tools should ideally replace some fundamental or repetitive tasks in the creative process, enhancing efficiency and saving unnecessary trouble, such as time spent on coloring and lighting. There is a general consensus among artists regarding the optimal AI tool: it should replace repetitive and basic tasks, while humans should still be responsible for thinking and incorporating inspiration. Some artists provided examples, suggesting that decisions like what to paint and finalizing the style should be made by humans, while intermediate processes, such as sketching and coloring, can be partially handled by AI. In more complex scenarios like video production, individual frames might require partial AI creation, and sound can also be partially AI-generated, but the final integration and processing should be human-led, encapsulating human thought, ideas, and creative intent.

Some artists adopt an overall stance that neither requires nor desires AI participation in their work. They believe that their current creative process still involves stages where AI has limited involvement. For example, an artist specializing in fashion design mentioned that while early stages of clothing design might utilize AI tools to some extent, the overall process remains predominantly hand-drawn. Moreover, the workload in these initial stages isn't particularly significant. In later stages, the artist focuses on physical collaging, material selection, and actual garment creation—tasks that are currently challenging to automate with AI. These stages require the artist's own reflection, observation, and action. Therefore, they express little concern about AI's future involvement, as they believe such participation is unlikely.

There are also many artists who adopt a resigned attitude. That is to say, although they dislike AI-generated works and the effects produced by AI, they acknowledge that the continuous evolution and advancement of AI are inevitable. These artists realize that they must gradually become familiar with using AI in their work and eventually treat it as a tool. Otherwise, they risk being replaced by AI in the commercial sector. For example, many large game companies have already laid off their art staff and switched to using AI for image production, which has had a significant impact on mid- to low-level artists. While high-end artists may remain difficult to replace in the short term, the overall trend of AI development is forcing more artists to accept this situation. Despite feeling conflicted about AI, these artists believe that, in the long run, they can learn to use AI and leverage AI tools to serve their needs. They acknowledge that AI can replace some of their repetitive tasks, such as quick image production or communication processes.

% 感觉是钱的问题

\subsection{Degradation of Aesthetic Value}
Some artists believe that the involvement of AI in the creative process may actually degrade the aesthetic standards of the artist community. These concerns are primarily divided into two aspects: the reduction of public aesthetic appreciation and the decline in artists' own aesthetic standards. 

Most artists think that the general public's aesthetic appreciation will diminish, although they believe their own aesthetic standards will remain relatively high. This is mainly because the general public has a limited ability to discern artistic quality and may perceive AI-generated works as aesthetically pleasing. However, many artists find these AI creations to be lacking in many areas, not only containing numerous flaws but also failing to meet aesthetic standards. If such works become widespread on social media and other platforms, they may lead to a gradual decline in the aesthetic standards of the current generation. 

Additionally, some respondents highlighted that this decline in aesthetic standards could significantly impact children. As children have not been exposed to many manually created works from before the advent of AI, their aesthetic sense is primarily shaped by AI-generated content prevalent today, potentially leading to a poor aesthetic sense. Finally, some respondents mentioned that their own aesthetic standards might also decline. This is because prolonged use of AI initially makes them find AI works unattractive, but through continuous interaction and mutual learning with AI, they might gradually become assimilated to AI aesthetics. This shift in artists' aesthetic perception could be risky, as it might hinder their ability to create works that are truly appreciated by society. Some artists refused to answer this question, arguing that aesthetic issues cannot be judged in this manner.

However, there are also some artists who thought the AI-generated content would not further degrade the aesthetic value of the public. This is because, even before the advent of AI technology, a large portion of the general public was already exposed to numerous mass-produced commercial artworks. The quality of these commercial images was not significantly higher than those generated by AI technology. Consequently, the overall impact on public aesthetic sensibilities remains substantial. While it is true that AI-generated images might also fall short of expectations, their use is unlikely to exacerbate the existing situation. For example, certain groups of men enjoy viewing comic images of women with attractive figures and revealing clothing. The production of these images is likely to utilize mass-generation techniques and does not necessarily require high artistic quality or meticulous craftsmanship. These images only need to satisfy visual and sensory stimulation or fulfill men's fantasies about women's bodies. This subset of commercial art, which caters to specific audiences, existed even before the advent of AI and has not undergone significant changes with its emergence.

Some participants mentioned that allowing the public to create AI-generated works or use AI tools themselves might not necessarily yield negative results. Many individuals would greatly enjoy the process of creating with AI, especially when generating their own AI avatars or producing desired images \cite{sanchez2023examining}. Although the aesthetic quality might not always be high, for them, it represents a deeply engaging way to participate in the creative process.

% 会让画师研究更多的风格，会让整体审美的取向更多
% 对消费者的欺骗，生成的是AI，可能不会和实物那么的贴合。不是AI技术问题，从卖家角度来看，AI生成的东西不会和实物完全一致。单纯让AI跑一个物品出来，很多买家会参照图片来购买。

% 纯插画和平面设计影响比较大，淘汰一大部分能力不是很强的创作者。对艺术家本身影响不大，所用的资源、人群和AI绘画面向的人群不完全重合，有自己独立的市场。

% 对一些专业影响不是很大，对信息设计和工业设计，对AI绘画的看法一般来说比较开放，他们把AI绘画当作一款工具，可以提升效率。
% 绘画、视觉传达对AI的接受程度不是很高。一部分因素，是被替代还是把AI当成工具来用。

% 有创造能力
% 不同人的工作流不太一样
% 带来一些新的东西，Ai实力的话，网上很多形式，作业和艺术创作基本上用不上。
% 小说推文，视频流经常能看到完全AI绘画完成的低质小游戏游戏界面，低端市场站稳脚跟
% 被使用者用来做不好的事情，换脸、造谣、违科普，没有办法筛查使用者的道德
% AI图随处可见，动态、朋友圈、微博、贴吧、小红书都能刷到Ai相关的广告
% 经常出现在大众视野里面，会造成大量审美疲劳，这是AI画的风格是好看的。画的图被广泛认为是好看的很恐怖。在一定程度上对于某些人提高了一些，把某些人审美封死在了AI的效果上。
% 让人觉得比较害怕，以后Ai都觉得，
% 缓解方法：非常大的变革，这种精细程度的画生成的快。贴近生活的各种信息和设计、UI设计、插画和动画，新型的产品。
% 更像人画的图 - 好看
% 普遍画的Ai的图没有达到这个标准，练他的人本身很多都是美术科班出身的，更像是一种工具，练出来什么样的作品，上限和下限很高。分都分不出来也练不出来
% 会用Ai学习程序相关的知识，是万能的老师。学习工具的使用，编程是实现想法的工具，需求AI可以给你，想实现的东西是一套一套的东西，喜欢自己写，把问题写出来
% 草稿、底色、成图，AI使用起来不太顺手，没有办法改线稿，每办法改线稿和底色，要重新画，局部重绘。人的工作流出线稿、底色、一版一版的出
% 肯定希望工具和自己更顺手一些
% 玩家对AI比较抗拒，画师的立绘，对Ai抵触，本身是画师
% 

\section{General Discussions and Design Implications}

\subsection{The Patterns of Human-AI Collaboration}
Different majors exhibited different collaboration patterns. For art majors like dyeing and weaving, the collaboration was more like using AI to facilitate coloring and drawing \cite{kim2022colorbo}. For sculpture majors, although there were literature about collaborative sculpture creation \cite{mingyuan2023cosculpt}, most participants believe the sculpture creating process should contain the thoughts of artists and could not be replaced or partly assisted by AI.

Notably, visual artists exhibited different collaboration patterns from designers and human-centric interaction researchers. The past literature illustrated the system facilitating co-creation in interaction design \cite{rezwana2023designing}. However, due to the aesthetic value of art works, these systems were not directly applicable to artists. The collaboration modes we found when artists collaborate with AI systems were still of the most basic manners. This also reflected the in-effectiveness of the previous literature \cite{choi2024creativeconnect} in applying to art practitioners' group due to learning cost and the lack of motivations. Participants echoed that the learning cost of stable diffusion, etc is low. However, mastering these tools was not easy.

\subsection{The Importance of Interviewing Novice Artists}

From the interview, some participants revealed that ``when an art worker has their own creating style, they no longer need to concern about their living issues.'' They also revealed that many artists would not choose AI-assisted creating simply because they lacked the motivation. These famous artists could sell their creative work with a high price without the potential competition of AI because they already have many faithful customers to pay. Even for those who were willing to use AI, they could easily profit much because of the higher visual aesthetics \cite{caramiaux2022explorers}. However, most of the novice artists have no established creative styles because they have less experience. From the interview, their concerns about whether the AI would replace their own work, whether AI could facilitate them to boost their working efficiency, whether AI would reduce their hourly wage \cite{shi2023understanding}, were greater \cite{wang2021differences}. Besides, the future market would increasingly be shaped by the opinions of these novice artists, with just the same cases as other works which focused on young adults when investigating the opinions towards newly innovated techniques \cite{mingyuan2023cosculpt}. Furthermore, compared with the professionals in the commercial industry who focused on specific production \cite{vimpari2023adapt}, the novice visual artists have greater potential to cultivate new collaboration pattern with AI and benefit more instead of being replaced. Their opinions were also more beneficial for designing artist G-AI collaboration. Different from the other groups such as online artists \cite{allred2023art} and designers \cite{kim2021tool}, our groups encountered more various art categories.

There were many disciplines that could potentially be affected by G-AI, such as musicians \cite{shi2023understanding}, designers \cite{wan2024felt}, etc. Their workflow and problems are not the same as visual artists. The problems of visual artists are not the same as novice visual artists. In fact, visual art is among the most affected by the G-AI, thus the opinions from novice visual artists, which were potentially among the most probable group to be replaced, is of high importance.

\subsection{Design Implications}

Numerous artists mentioned the potential hinders towards human G-AI collaboration: 1) the AI tool was hard to master and in most cases it could not understand the viewpoints of art workers. 2) the art workers could not explicitly explain their opinions and thoughts simply through voice. 3) the creative process is different from the canonical and natural creating process of art workers, which refined the paintings several times and split the drawing to different sessions. In comparison, AI completed the drawing in a short time. 4) the quality of AI-assisted work is fluctuating. In some cases the drawings were satisfying but in most cases it is unusable. It needs artists with great effort in identifying the seemingly homogeneous work. Thus, this work proposed several design implications to facilitate human AI collaboration:

\noindent \textbf{Designing Interfaces to Reduce the Learning Cost of AI Tools.} For artists, it is unnatural to transfer thoughts into natural language and describe the scene using natural language. Most artists choose to feed outline sketching into the AI tools to better guide generation. It is recommended that more GUI-based and drawing guided tools should be developed for artists \cite{zhang2023generative}. For example, the artist could draw lines in the picture or use some basic color to guide the painting of AI.

\noindent \textbf{Creating More Natural Image Generation through Modifying the Generation Process.} The current basic generation process is guided by diffusion process, which is the de-noising process in the core \cite{ho2020denoising}. However, the perception of the artists was that the AI generated all the content at last in a short time, without painting each part in the middle. This made the integration of intermediate thoughts hard. Besides, artists could only reverse the process to modify a well-completed painting if they were unsatisfied about the content and the structure. This is unnatural for participants. Thus, the future image generation should consider the canonical steps and support the splitting of these steps. For each step, the painting result should be presented as an intermediate layer. At the end, the AI tools should also refrain from drawing a well-completed painting. Instead, it should place the highlight on some focus object in the painting and ``omit'' the others.

\noindent \textbf{Stabilize the Quality of Drawing} Given the massive selection and grading of artists' regarding the aesthetic value of the drawings, the AI tools should stabilize the quality of drawing. This could be achieved through mimicking or learning the potential patterns of high-quality paintings through a single artist's feedback, or through collaboratively learning numerous artists' work together. This process could improve the experience, reduce the selection cost of artists and at the same time induce no additional ethical or copyright issues \cite{dee2018examining,chiou2022copyright}.

However, it should also be noticed from the interview and the past work that human AI collaboration not necessarily lead to higher collaborative performance \cite{saffiotti2020human,hitsuwari2023does}. Most art practitioners agreed that the collaborative performance equals the artist's own aesthetic value \cite{chang2021towards} and some artists also echoed that using AI would decrease their own efficiency. These proved that prompting AI is not easy.

\section{Ethical Considerations}

We strictly adhered to the guidelines outlined in the Menlo Report \cite{bailey2012menlo} and the Belmont Report \cite{beauchamp2008belmont} in conducting our experiments. The experiment was approved by the Institutional Review Board (IRB) of our campus. In designing the experiments, we made every effort to avoid asking questions that could be offensive or potentially offensive to participants, such as whether they believe their drawing skills are superior or inferior to AI. Prior to the experiment, we briefed participants on the main content and purpose of the study and obtained their informed consent. We provided appropriate explanations to participants who had concerns about the experimental content. Participants were informed that they could withdraw from the experiment at any time if they had any objections or dissatisfaction with the experimental content or the informed consent process. We also informed participants that they could choose not to answer any questions and that participation could be terminated at any stage, all on a voluntary basis. Participants received a compensation of 100 RMB upon completing the experiment. Furthermore, all participant information was anonymized. Audio recordings and transcriptions of the interviews were securely stored and effectively encrypted on a local machine. The study aimed at improving the collaboration correlation between artists and the AI tools, and participants also reflected that they would acquire new viewpoints, inspirations and benefit towards integrating AI into their creative work during the communication.

\section{Conclusion}
In this work, we initiated the study on novice artists' perception towards AI generated creative work. Through an in-depth interview towards novice artists (N=20), we identified the different patterns, opinions and practice of four different aspects: 1) the evolving opinions of novice artists, 2) the practice of novice artists , 3) the stakeholders' opinions, 4) the future expectations. We detailed several design implications to facilitate the co-creation of artists and AI including: 1) designing learning and expression interfaces with lower learning cost, 2) designing natural creation process which involved both participants and AI, 3) stabilizing the drawing quality through supervised learning. This work served as the first step towards understanding the human-AI co-creation practice.

\bibliographystyle{ACM-Reference-Format}
\bibliography{sample-base}

% 版权，版权的解决方案
% AI没有意识，人有意识，人会主动画不同的东西
% 生成的图很普遍很大众化的风格
% 米山舞

\appendix
\section{The Interview Outline}\label{sec:outline}

The following are questions in the outline. Experimenters would append questions if the participants' answers were of specific interest to them. 

\begin{enumerate}
    \item What is your major?
    \item Do you have any specialized courses this semester or last semester? What professional experiences or skills do you have in your field?
    \item What do you consider to be the main components of your professional creative process? How do you complete each component?
    \item Briefly describe some of your experiences with using AI to assist or participate in your creative work.
    \item What are your attitudes and opinions towards AI?
    \item How many years have you studied drawing (not self-taught, but under formal instruction)?
    \item How many times per week do you use AI drawing or similar art-assistance tools?
    \item Which parts of your professional creative process do you think could potentially be replaced by AI?
    \item Which parts of your regular drawing or sketching activities do you think could be replaced by AI?
    \item Which parts do you believe cannot be replaced by AI, and why?
    \item Do you think AI possesses creativity and innovation? Why or why not?
    \item What do you consider to be the greatest strengths and weaknesses of AI?
    \item Why are you willing or unwilling to use AI in your creative process?
    \item What are the opinions of those around you regarding AI?
    \item Why do they hold such attitudes and opinions about AI?
    \item Have you worked on any projects? What are the clients' views on AI, and what is their attitude towards your use of AI?
    \item Why do you think the clients have such opinions?
    \item What do you think the potential audience's views are on AI, and what is their attitude towards the use of AI in artworks?
    \item Do you think the trend of AI development is irreversible? Why or why not?
    \item What impact do you think the widespread dissemination of AI-generated artworks will have on the audience?
    \item Do you think your learning of AI is more self-taught or through formal education?
    \item Do you find learning and using AI difficult? What is the most challenging part?
    \item In your opinion, how does AI generate artworks?
    \item What do you think your future relationship with AI will be?
    \item Which parts of your work do you think AI might help you replace, and what tasks cannot be replaced?
    \item Do you believe you can clearly express your creative needs to AI? Why or why not?
    \item Do you think the quality of AI-generated images you produce (which may be unsatisfactory) is due more to the technology itself or your usage?
    \item What is your understanding of the copyright issues related to AI creation, and would you be willing to use your works for AI training?
    \item Have you used your own works for training AI? How was the result? How does AI-generated artwork that matches your style affect your creative process?
    \item Do you have any other questions for us?
\end{enumerate}

\end{document}